\documentclass[review,12pt]{elsarticle}
\usepackage{graphicx}

\journal{Physics C}

\begin{document}

\begin{frontmatter}

\title{Global and local critical current density in
    superconducting SmFeAsO$_{1-x}$F$_x$ measured by two
    methods}

\author[kyutech]{E.S. Otabe\corref{cor1}}
\ead{otabe@cse.kyutech.ac.jp}
\cortext[cor1]{Author to whom correspondence should be addressed. }

\author[kyutech]{M. Kiuchi}
\author[kyutech]{S. Kawai}

\author[fit]{Y. Morita}
\author[fit]{J. Ge}
\author[fit]{B. Ni}

\author[CAS]{Z. Gao}
\author[CAS]{L. Wang}
\author[CAS]{Y. Qi}
\author[CAS]{X. Zhang}
\author[CAS]{Y. Ma}

\address[kyutech]{
Faculty of Computer Science and Systems Engineering,
    Kyushu Institute of Technology, 680-4 Kawazu, Iizuka,
    Fukuoka 820-8502,
    Japan}

\address[fit]{
Department of Life, Environment and Material
    Science, Fukuoka Institute of Technology, 
    3-30-1 Wajirohigashi, Higashi-ku, Fukuoka 811-0295, Japan}

\address[CAS]{Key Laboratory of Applied Superconductivity,
  Institute of Electrical Engineering, Chinese Academy of
  Sciences, PO Box 2703, Beijing 100190, People's Republic
  of China
}

\begin{abstract}
The critical current densities of polycrystalline bulk
SmFeAsO$_{1-x}$F$_x$ prepared by the powder-in-tube (PIT)
method and by a conventional solid-state reaction were
investigated using the remnant magnetic moment method and
Campbell's method. Two types of shielding current,
corresponding to global and local critical current densities
$J_{\rm c}$ were observed using both measurement methods.
The global and local $J_{\rm c}$ were on the order of
$10^7$~A/m$^2$ and $10^{10}$~A/m$^2$ at 5~K, respectively.
The local $J_\mathrm{c}$ decreased slightly with increasing
magnetic field. The global $J_{\rm c}$ was independent of
the preparation method, while the local $J_{\rm c}$ was
larger for samples prepared by PIT than for those prepared
by solid-state reaction.

\end{abstract}


\end{frontmatter}

% -------------------------------------------------------

\newpage

\section{Introduction}

The recent discovery of superconductivity in Fe-based
compounds has attracted considerable attention, since the
superconducting mechanism is different from that of Cu oxide
superconductors, and Fe-based superconductors are expected
to be useful in a wide array of future applications 
\cite{kamihara}. The critical temperature can reportedly be
enhanced over 50~K by incorporating another rare earth
element \cite{ren}-\cite{eisaki}. Although this value is
still lower than the boiling point of liquid nitrogen
(77.3~K), further discovery can not be denied. In addition,
it has been reported that the upper critical field may be
higher than 100~T \cite{hunte, senatore}.  Therefore,
Fe-based superconductors may allow applications with higher
operating temperatures.

We previously reported the preparation of
SmFeAsO$_{1-x}$F$_x$. Polycrystalline bulk specimens were
synthesized by conventional solid-state reaction \cite{wang1}
and by the powder-in-tube (PIT) method \cite{gao1}.  Recently
a one-step synthesis was reported instead of the usual
two-step synthesis \cite{ma1}.

Since the specimens were polycrystalline, it was expected
that two types of shielding currents should be observed in
the magnetization measurements \cite{yamamoto}. They are
known as local and global critical current densities. That
is, connection between grains is too week and the difference
of shielding current inside grains and through the grains are
large as two to three order of magnitude. They are also
called intra- and inter-grain shielding currents. However,
the loop size of shielding current is not directly
corresponding to the size of the grain for the case of very
low quality polycrystalline specimen, since the shielding
current may flow several grains or limited area of the
grain. Therefore, in the present paper, the local and global
critical current densities are used instead of intra- and
inter-grain critical current densities.

In our previous works \cite{ma1}, the local and global critical
current densities are roughly estimated from the simple
magnetization measurement. However, according to this
method, the accurate measurement can not be expected since
the measured magnetic moment is a type of average of
the local and the global critical current
densities \cite{nee}. Therefore, an accurate measurement for the
local and global critical current densities is required. 

In reference \cite{yamamoto}, the two kinds of shielding
currents were estimated from the remnant magnetization as a
function of maximum applied field, and it was explained that
these currents corresponded to the local and global critical
current densities. Although this measurement method is
simple, the external magnetic field is restricted to zero,
so it is impossible to obtain the magnetic field dependence
of the critical current densities. In contrast, the two
kinds of critical current density can be measured separately
by the use of Campbell's method \cite{nee, campbell}. This
method can much more effectively investigate the critical
current densities in bulk specimens which have a complicated
current path. Moreover, there are few reports on the
characteristics of the critical current density of Fe-based
superconductors. 

In this paper, the two kinds of critical current density
were measured using both the remnant magnetization method
and Campbell's method, with specimens prepared by two
different preparation methods.  The difference in values
obtained by the two measurements is discussed.

%--------------------------------
\section{Experimental}

The sample specifications are listed in Table~1.  Specimen
\#2 was prepared by the PIT method \cite{gao1}, and the other
two specimens \#3, \#4 were prepared by conventional
solid-state reaction with different nominal compositions of
fluorine \cite{wang1}. The details of their preparation are
described in the references.  The critical temperatures of
the specimens, measured from the temperature dependence of
magnetic moment in field cool and zero field cool, were
about 50~K.

The structure of the specimens were examined by scanning
electron microscopy (SEM). It was found that the typical
grain size was about the same as was reported previously,
i.e., in the range of 1--20~$\mu$m \cite{wang1, gao1}. There
exists unreacted particles which size is smaller than the
grain size. Although the connection between grains is almost
insufficient, strong connected grains are also
observed. That is, the size of loop shielding current is
widely distributed. Therefore, the size of the loop is
assumed as be 10~$\mu$m in the present study.

The remnant magnetic moment $m_{\rm R}$ at zero magnetic
field was measured after an excursion up to an external
maximum magnetic field $H_{\rm a}$. The shield current,
which is identified as the critical current density $J_{\rm
  c}$, is related to $H_{\rm a}$. Assuming a
superconducting slab of thickness $w$ and that the external
magnetic field is applied in parallel to the wide
surface, the remnant magnetization $M_{\rm R}$ is given as
\begin{equation}
  \label{eq:1}
  {\def\arraystretch{2.5}
  \begin{array}{lll}
    M_{\rm R}  & =\displaystyle {H_{\rm a}^2\over 4H_{\rm p}}; & H_{\rm a}
    < H_{\rm p} \\
    & =\displaystyle -{H_{\rm a}^2 \over 4 H_{\rm p}}+H_{\rm a} -{H_{\rm
        p}\over 2 }; & H_{\rm p} <H_{\rm a}< 2 H_{\rm p} \\
    & =\displaystyle {H_{\rm p} \over 2}; & H_{\rm a}>2 H_{\rm p}
  \end{array}}
\end{equation}
where $H_{\rm p}$ is the penetration depth which is equal to
$J_{\rm c} w/2$ based on Bean's model. Therefore, the
derivative of Eq. (\ref{eq:1}) is given by
\begin{equation}
  \label{eq:2}
  {\def\arraystretch{2.5}
  \begin{array}{lll}
    \displaystyle {{\rm d} M_{\rm R}\over {\rm d} H_{\rm a}} 
     & =\displaystyle {H_{\rm a}\over 2H_{\rm p}}; & H_{\rm a}
    < H_{\rm p} \\
    & =\displaystyle -{H_{\rm a} \over 2 H_{\rm p}}+1; 
    & H_{\rm p} <H_{\rm a}< 2 H_{\rm p} \\
    & =\displaystyle 0; & H_{\rm a}>2 H_{\rm p}.
  \end{array}}
\end{equation}
Hence, ${\rm d} M_{\rm R}/{\rm d} H_{\rm a}$ shows a peak at
$H_{\rm a}=H_{\rm p}$, where $J_{\rm c}$ can be estimated
from the measured $H_{\rm p}$. In the present study, remnant
magnetic moment $m_{\rm R}$ as a function of maximum
magnetic field $H_{\rm a}$ was measured using a SQUID
magnetometer.

Campbell's method was used to measure the critical current
density of the specimens. The DC and superimposed AC
magnetic fields were applied to the specimen. The frequency
of the AC magnetic field was 97~Hz, and the maximum AC
magnetic field was 10~mT. The total magnetic flux $\Phi$
penetrating the specimen was calculated from the signals of
pick-up and cancel coils as a function of the AC magnetic
field amplitude $b_{\rm ac}$.  Then the penetration depth of
the AC magnetic field $\lambda'$ was derived as
\begin{equation}
  \label{eq:3}
  \lambda' = {1\over 2 w}{\partial \Phi \over \partial
    b_{\rm ac}}.
\end{equation}
The slope of the $\lambda'$-$b_{\rm ac}$ plot gave $1/\mu_0 J_{\rm
c}$. The value of $\lambda'$ became saturated to a
constant representing the center of the superconductor at a
large $b_{\rm ac}$ value. If there were two kinds of magnetic
moments in the specimen, two slopes would be observed and two
different critical current densities could be
evaluated. Temperature of specimen was controlled by
cryocooler above 18~K.

%--------------------------------------

\section{Results and Discussion}

Fig.~1(a) shows the external applied magnetic field
dependence of the derivative of the magnetic moment of
specimen \#2 at various temperatures. Two peaks were clearly
observed.  The same behavior was observed for specimen \#3,
which was prepared by solid-state reaction unlike the PIT
preparation of \#2. The first peak at smaller $H_{\rm a}$ is
related to the global critical current density, since the
peak rapidly shifted to a lower magnetic field with
increasing temperature, and the estimated value of $J_{\rm
  c}$ was much smaller than the local critical current
density. It was reported that the global critical current
density measured by four probe method reached zero at
temperatures higher than 20~K \cite{gao1}. In contrast, the
second peak at larger $H_{\rm a}$ is related to the local
critical current density. It was found that the temperature
dependence of the local critical current density was smaller
than that of the global critical current density.

On the contrary, only one peak was found for specimen \#4,
as shown in Fig.~1(b). This peak corresponds to the local
critical current density. The global critical current
density of specimen \#4 was too small to be measured under
the present conditions.

The global and local $J_{\rm c}$ evaluated from the peak
positions of the derivatives of $m_{\rm R}$ are shown in
Fig.~2. Since only one peak was found for specimen \#4,
the global $J_{\rm c}$ was omitted. The value of the global
$J_{\rm c}$ was on the order of $10^7$~A/m$^2$ at 5~K and
could not be estimated for temperatures above 20~K, since
the peak in the derivative of $m_{\rm R}$ was too
small. This value agrees with that reported by Yamamoto {\it
  et. al.} \cite{yamamoto} The temperature dependence of
global $J_{\rm c}$ in specimens \#2 and \#3 were the
same. Therefore, the global $J_{\rm c}$ was too small and
was significantly affected by voids and weak links between
grains in the polycrystalline bulk specimens, both of which
are known to exist in Cu oxide superconductors with poor
properties.  It is well-known that the temperature and
magnetic field dependences of $J_\mathrm{c}$ affected by
weak links are large.

In contrast, the local $J_{\rm c}$ at 5~K was over
$10^{10}$~A/m$^2$, 1000 times larger than the global $J_{\rm
  c}$, which was consistent with the results estimated in
ref. \cite{yamamoto}. It was found that the local $J_{\rm c}$
was quite dependent upon the method of synthesis, as shown
in Fig.~2(b). Moreover, the temperature dependence of
local $J_{\rm c}$ was also different. That is, the local
$J_{\rm c}$ of specimen \#2 prepared by PIT was higher than
specimens \#3 and \#4 prepared by conventional solid-state
reaction. Therefore, it is reasonable to expect that the
characteristics of the critical current density could be
improved by further developing the preparation method.

Fig.~3 shows an example of $\lambda'$ as a function of
$b_\mathrm{ac}$ measured using Campbell's method. There were
two slopes, corresponding to very-low and high
$b_\mathrm{ac}$.  According to Campbell's method, the first
and second slopes are related to the global and local
$J_{\rm c}$, respectively.  It was found that the global
$J_{\rm c}$ was only observed at zero DC magnetic field by
Campbell's method. That is, the global $J_{\rm c}$ was very
small, and the magnetic field dependence of $J_\mathrm{c}$
was extremely poor, as is frequently observed as weak link
property in Cu oxide superconductors \cite{matsushita1991}.
This result agrees with that obtained by measurement of the
remnant magnetic moment.

The temperature dependence of the local $J_\mathrm{c}$ for
specimen \#2 at zero DC magnetic field estimated using
Campbell's method is also shown in Fig.~2(b). Although
this value of local $J_\mathrm{c}$ was quantitatively
different from that obtained by the remnant magnetic moment
method, the order of magnitude was the same. The reason for
the quantitative difference is mainly caused by the
difference of measurement methods. For example, although the
change of the superconducting volume fraction does not
affect to $J_\mathrm{c}$ by the remnant magnetic moment
method, it causes large affect to Campbell's method. In
addition, the local fluctuation of $J_\mathrm{c}$ can be
measured by Campbell's method, while the measured
$J_\mathrm{c}$ is an average value by the
remnant magnetic moment method.

Fig.~4 shows the magnetic field dependence of local
$J_{\rm c}$ at various temperatures, as measured using
Campbell's method. The values of local $J_\mathrm{c}$ ranged
around $10^9$~A/m$^2$, consistent with the prediction based
on the remnant magnetic moment method. The local
$J_\mathrm{c}$ slightly decreased with increasing magnetic
field and temperature, except in the temperature region near
$T_\mathrm{c}$, where a peak effect appeared in the magnetic
field dependence of local $J_\mathrm{c}$.

To investigate the value of $J_{\rm c}$ over a wide range of
magnetic fields, $J_\mathrm{c}$ was estimated from the
hysteresis of magnetic moment $\Delta m$ at various
temperatures, measured using a SQUID magnetometer as shown
in Fig.~5.  This $J_\mathrm{c}$ is a type of average of
the local and the global $J_\mathrm{c}$, and is
different from both local and global $J_\mathrm{c}$ \cite{nee}.  The
peak effect was widely observed at various temperatures. The
first peak field was at about 0.1~T and was consistent with
the result from Campbell's method, as shown in Fig.~4.  It
is known that the peak effect is caused by the transition of
the vortex lattice in Cu oxide
superconductors \cite{matsushita2001}-\cite{henderson}.
However, the reason for the first peak effect in present
case is unknown.  

On the other hand, the second peak field was observed at
magnetic fields of over 1~T. This seems unrealistic, since
$J_\mathrm{c}$ was observed at 50~K, which is higher than
the critical temperature of specimen \#3.  The results of
second peak effect and observed $J_\mathrm{c}$ at 50~K could
be ascribed to the possible existence of unreacted iron in
the specimens. Since iron is ferromagnetic, the hysteresis
of magnetic moment due to the ferromagnetism would remain,
even at temperatures higher than $T_\mathrm{c}$, and would
disappear at magnetic fields larger than 1~T.  This was
confirmed by the temperature dependence of the magnetic
moment $m$ in field cool, in which $m$ showed a positive
value, even in the superconducting state.  These different
effects of residual ferromagnetic iron are the origin of the
observed difference in local $J_\mathrm{c}$ between the two
measurement methods, as shown in Fig.~2(b). Therefore, it is
necessary to accurately estimate the inference of
ferromagnetic property in the specimens.

%-------------------------------
\section{Conclusion}

In this work, two preparation methods were used to prepare
specimens of polycrystalline bulk SmFeAsO$_{1-x}$F$_x$.  The
global and local shielding current densities, which
correspond to the global and local critical current density,
respectively, were measured using the remnant magnetic
moment and Campbell's methods. The global critical current
density was very low, on the order of $10^7$~A/m$^2$ at 5~K,
and this value was independent of the preparation
methods. This is likely due to the effects of voids and weak
links between grains in the polycrystalline bulk
specimens. In contrast, the local $J_{\rm c}$ was about 1000
times larger than the global $J_{\rm c}$, and the value and
temperature dependence of $J_\mathrm{c}$ was largely
dependent upon the preparation methods. The estimated
$J_\mathrm{c}$ from the two measurements were in agreement.
Our results show that remnant unreacted iron in the specimens
significantly affected the results of the $J_\mathrm{c}$
measurements.

%------------------------------
\section*{Acknowledgment}

This research was supported by the Japan Science and Technology
Agency (JST), Transformative Research-project on Iron
Pnictides (TRIP), 2008. 

%-------------------------------------------------------

\newpage

Table 1 Sample specifications

\begin{tabular}{llll}
\hline
specimen No. & specimen & synthesis & critical temperature [K] \\
\hline
\#2 & SmFeAsO$_{0.7}$F$_{0.3}$ & PIT(powder in tube) & 50.4 \\
\#3 & SmFeAsO$_{0.7}$F$_{0.3}$ & solid state reaction & 48.5 \\
\#4 & SmFeAsO$_{0.6}$F$_{0.4}$ & solid state reaction & 51.1 \\
\hline
\end{tabular}

\newpage

%------------------------------------------------------

\section*{Figure captions}

\begin{itemize}
\item[Figure 1] Derivative of remnant magnetic moment as a
  function of maximum external applied magnetic field at
  various temperatures in (a) specimen \#2 and (b) specimen \#4.

\item[Figure 2] (a) Global and (b) local critical current
  densities as a function of temperature at zero magnetic
  field, estimated from remnant magnetic moment. Filled
  squares denote the results obtained using Campbell's method.

\item[Figure 3] Example of penetration depth of the AC
  magnetic field as a function of AC magnetic field by
  Campbell's method. The inset shows an enlarged plot of the small
  AC magnetic field region.

\item[Figure 4] Magnetic field dependence of local critical
  current density of specimen \#2 at various temperatures
  obtained using Campbell's method.

\item[Figure 5] Magnetic field dependence of the critical
  current density, directly evaluated from the hysteresis of
  the magnetic moment of specimen \#3 at various temperatures.

\end{itemize}

\begin{center}
\includegraphics[width=0.6\linewidth]{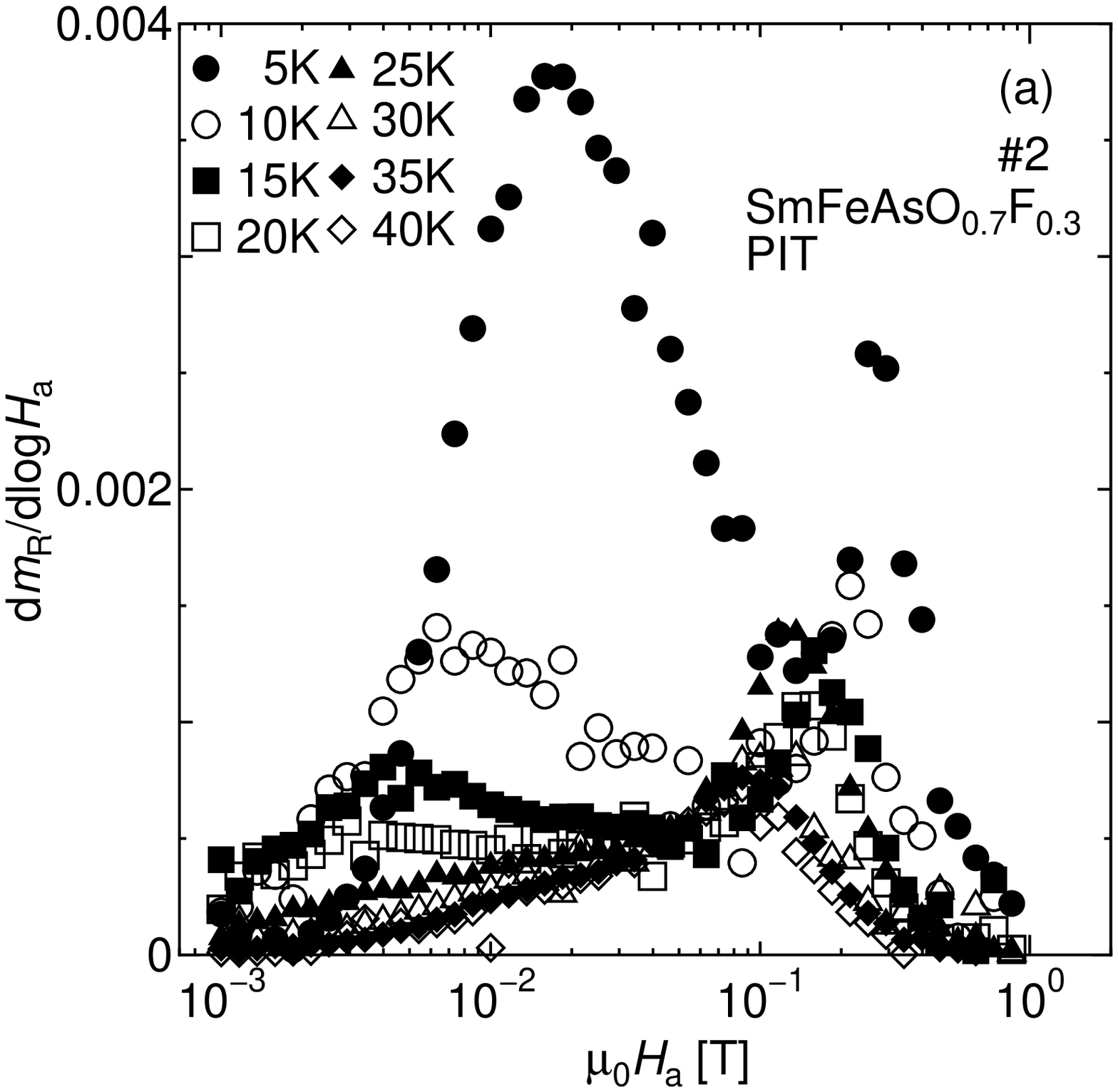}\\
\ \\
\includegraphics[width=0.6\linewidth]{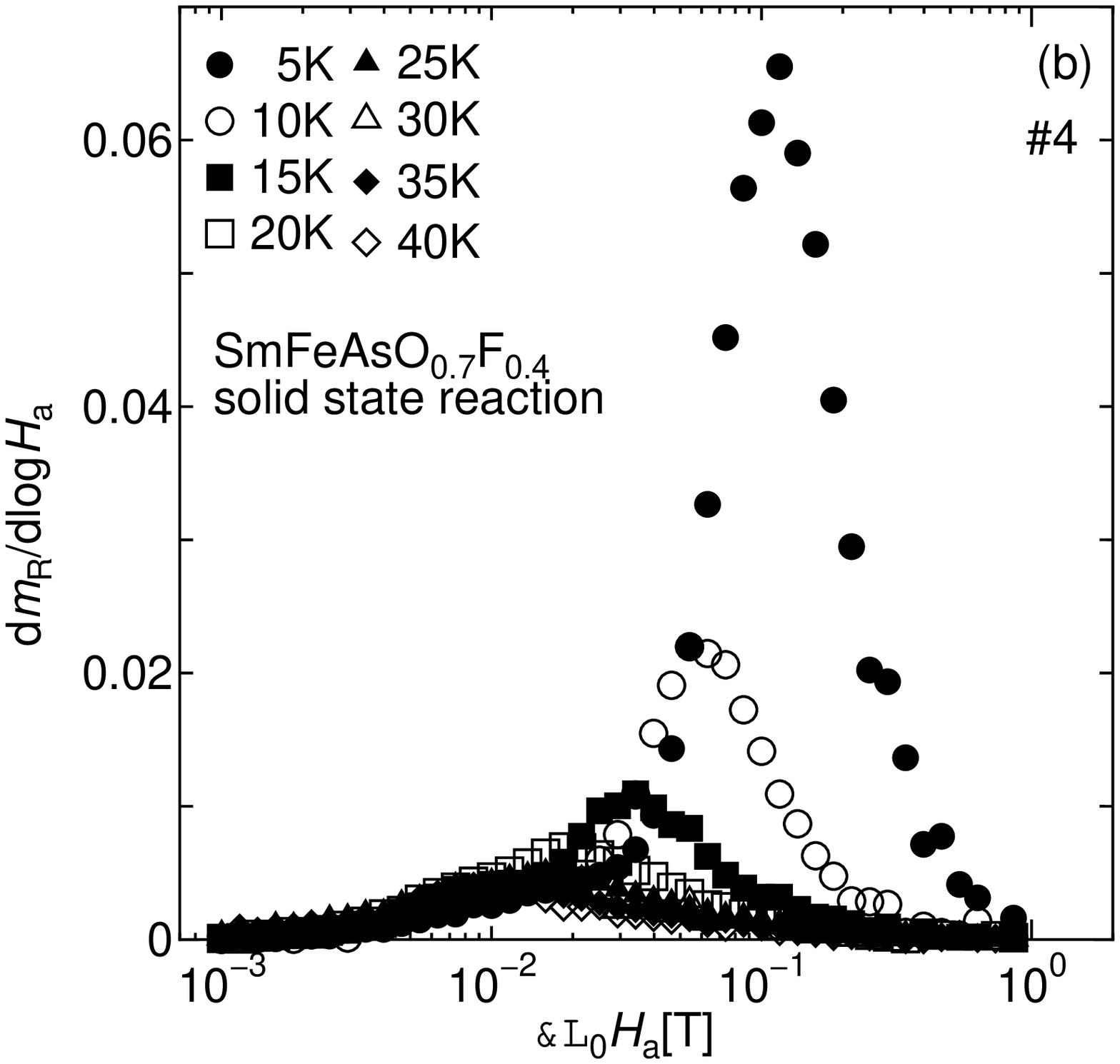}
\end{center}

\vfill
Fig. 1: E.S. Otabe {\it et al.\/} 
\newpage

\begin{center}
\includegraphics[width=0.6\linewidth]{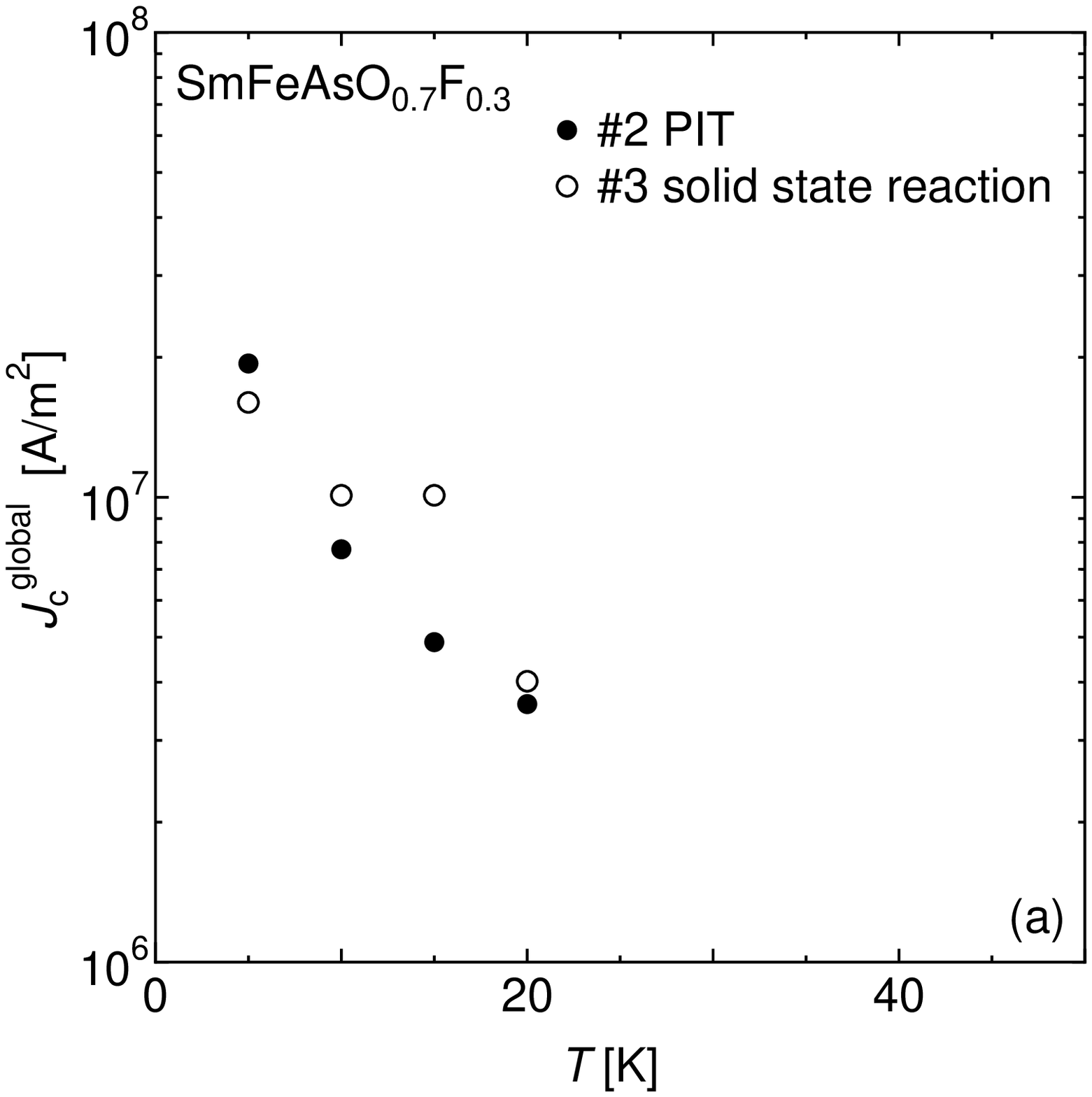}\\
\ \\
\includegraphics[width=0.6\linewidth]{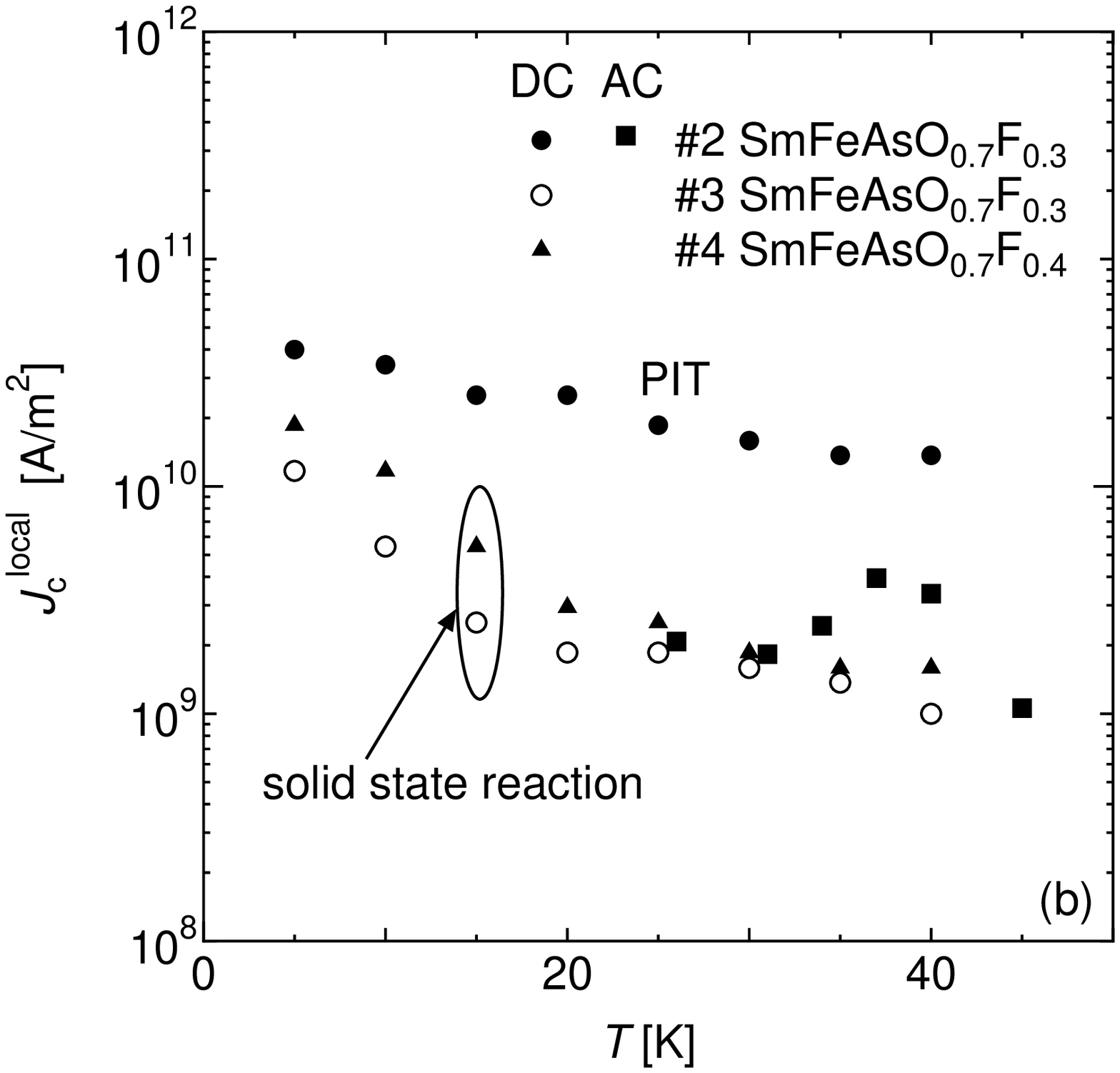}
\end{center}

\vfill
Fig. 2: E.S. Otabe {\it et al.\/} 
\newpage

\begin{center}
\includegraphics[width=\linewidth]{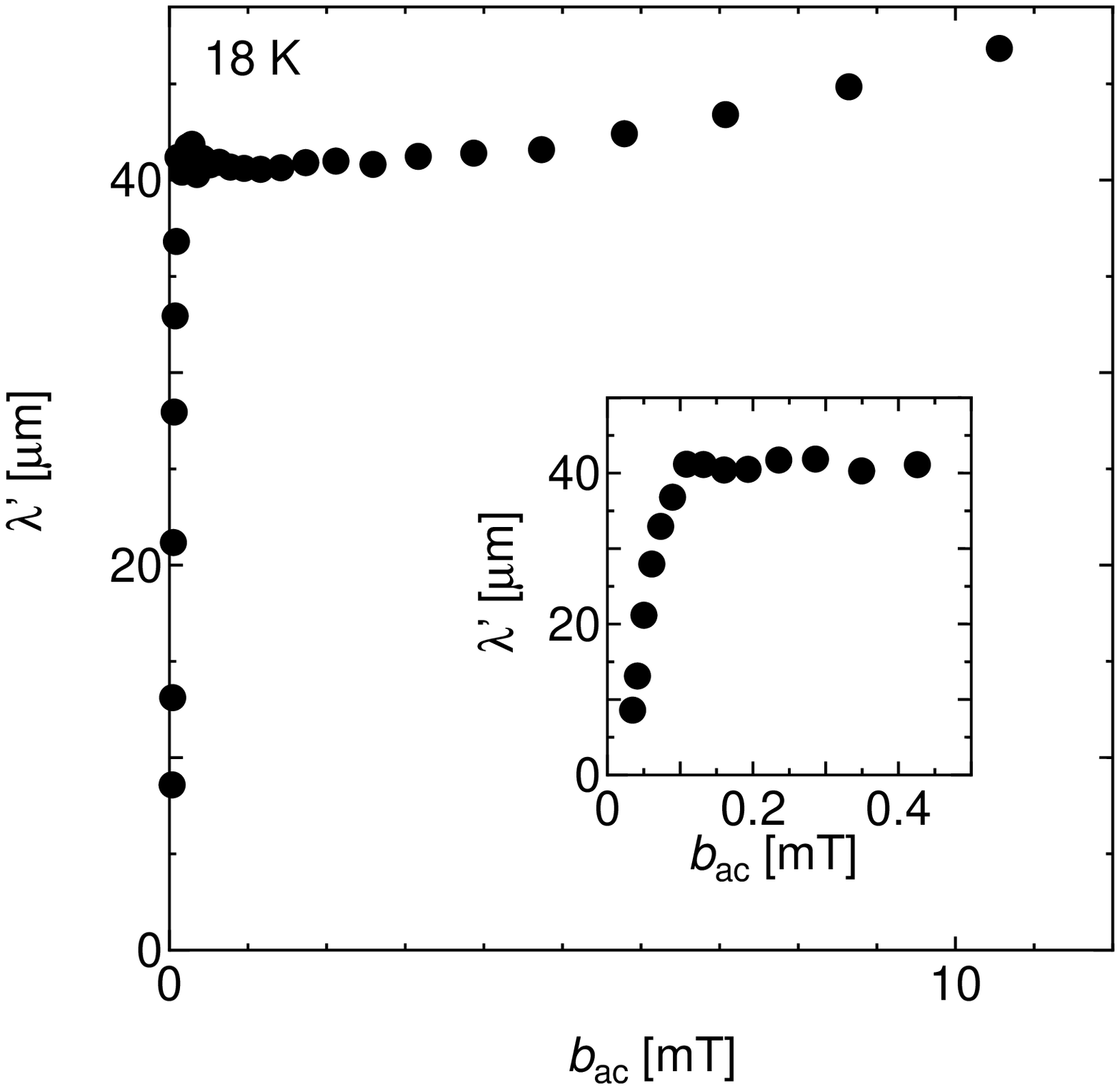}
\end{center}

\vfill
Fig. 3: E.S. Otabe {\it et al.\/} 
\newpage

\begin{center}
\includegraphics[width=\linewidth]{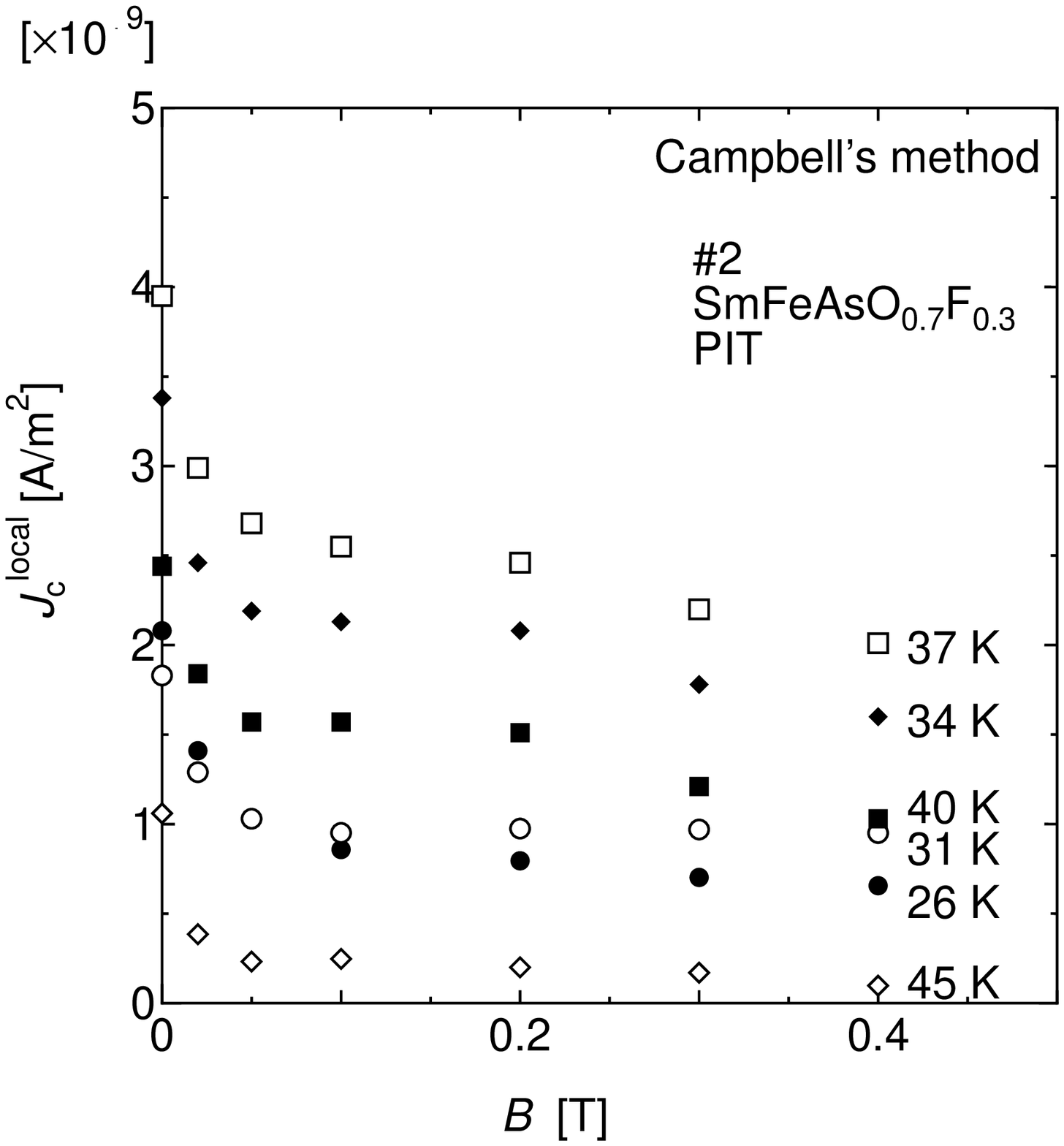}
\end{center}

\vfill
Fig. 4: E.S. Otabe {\it et al.\/} 
\newpage

\begin{center}
\includegraphics[width=\linewidth]{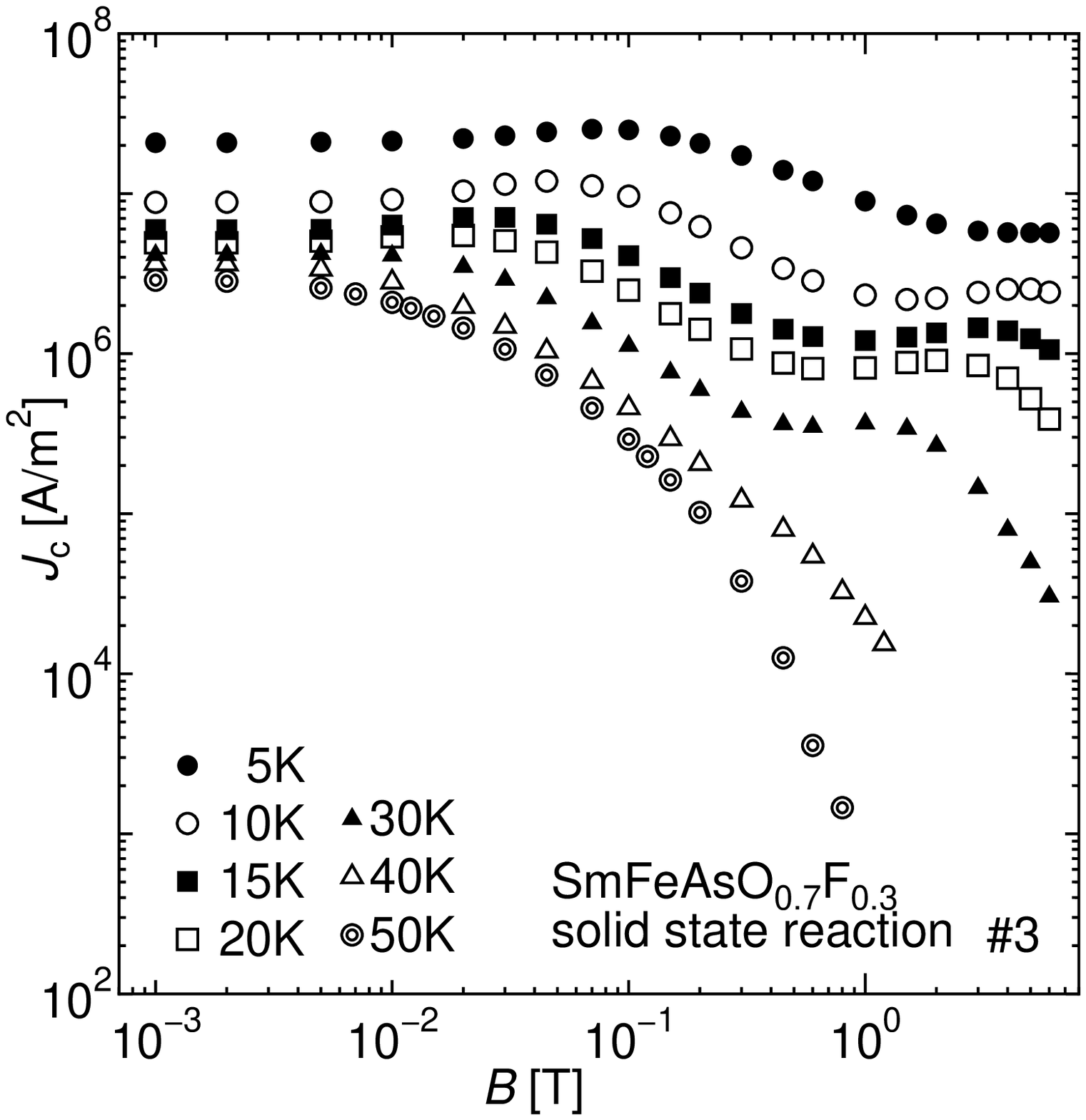}
\end{center}

\vfill
Fig. 5: E.S. Otabe {\it et al.\/} 
\newpage

\end{document}